\documentclass[journal]{IEEEtran}
\ifCLASSINFOpdf
\else
\fi
\usepackage{amsthm}
\usepackage{stfloats}
\usepackage{graphicx}
\usepackage{booktabs}
\usepackage{float}
\usepackage{caption}
\usepackage{amssymb,amsfonts}
\usepackage{listings}
\usepackage{url}
\usepackage{algorithm,algorithmicx}
\usepackage{algpseudocode}
\usepackage{enumitem}
\usepackage{mathrsfs}
\usepackage{amsmath}

\theoremstyle{plain}

\theoremstyle{definition}

\theoremstyle{remark}

\begin{document}
%
\title{Convexity Analysis of Optimization Framework of Attitude Determination from Vector Observations}
%
%
%

\author{Jin~Wu,~\IEEEmembership{Member,~IEEE,}
            Zebo~Zhou
            and~Min~Song
\thanks{This work was supported by National Natural Science Foundation of China under the grant of No. 41604025 and was also supported by State Key Laboratory of Geodesy and Earth’s Dynamics (Institute of Geodesy and Geophysics, Chinese Academy of Sciences) Grant No. SKLGED2018-3-2- E. (Corresponding author: Zebo Zhou)}
\thanks{J. Wu is with School of Aeronautics and Astronautics and School of Automation, University of Electronic Science and Technology of China, Chengdu, 611731, China. e-mail: jin\_wu\_uestc@hotmail.com.}
\thanks{Z. Zhou is with School of Aeronautics and Astronautics and School of Automation, University of Electronic Science and Technology of China, Chengdu, 611731, China. e-mail: klinsmann.zhou@gmail.com.}
\thanks{M. Song is with State Key Laboratory of Geodesy and Earth's Dynamics, Institute of Geodesy and Geophysics, Chinese Academy of Sciences, Wuhan 430077, China. e-mail: msong82@hotmail.com.}
}

\maketitle

\begin{abstract}
In the past several years, there have been several representative attitude determination methods developed using derivative-based optimization algorithms. Optimization techniques e.g. gradient-descent algorithm (GDA), Gauss-Newton algorithm (GNA), Levenberg-Marquadt algorithm (LMA) suffer from local optimum in real engineering practices. A brief discussion on the convexity of this problem is presented recently \cite{Ahmed2012} stating that the problem is neither convex nor concave. In this paper, we give analytic proofs on this problem. The results reveal that the target loss function is convex in the common practice of quaternion normalization, which leads to non-existence of local optimum.
\end{abstract}


%
\IEEEpeerreviewmaketitle

\section{Introduction}
%
%
%
%
\IEEEPARstart{A}{ttitude} determination has been extensively employed in mechatronic platforms and consumer electronics for orientation measurement \cite{Zhou2016, Zhou2013}. From vector sensor outputs, one can compute the optimal attitude transformation matrix accordingly.\\
\indent In the past 50 years, attitude determination from vector observations has been systematically studied. One of the most famous centers of such research is called the Wahba's problem posed by G. Wahba in 1965 \cite{Wahba1965} targeting to find out the least-square alignment of two point sets. The optimal correspondence of this problem had not been solved very effectively in later several years until the invention of Davenport's approach i.e. the q-method in 1968 \cite{Davenport1968}. The q-method converts the Wahba's optimization into an eigenvalue-seeking problem of the Davenport $\bf{K}$ matrix. In later research, the endeavors paid their most attention into finding efficient computation procedure of the characteristic polynomial to $\bf{K}$. This in fact generates a large variety of algorithms including famous ones e.g. the QUAternion ESTimator (QUEST, \cite{Shuster1981}), Fast Optimal Attitude Matrix (FOAM, \cite{Markley1993}), EStimator Of Quaternion (ESOQ, \cite{Mortari1997}) and etc. Our recent contribution called the Fast Linear Attitude Estimator (FLAE, \cite{Wu2018}) can also be categorized into this kind of solvers.\\
\indent The Wahba's solutions are especially applied to aerospace engineering for 3-axis satellite attitude determination where the sun sensor, nadir sensor, star tracker, magnetometer and gravimeter are invoked for vector observation outputs \cite{Crassidis2007}. In consumer electronics, where the sensor precisions are relatively low, the vector sensors e.g. accelerometer and magnetometer are usually integrated with gyroscope for more smooth estimates \cite{Yun2006}. For these estimators, there is always a need of measurement source for direct attitude reconstructions from sensors. According to bare computation resources, many algorithms are also developed to extract orientation by means of simple optimization methods. The Gauss-Newton algorithm (GNA, \cite{Marins2001}) is almost the first one doing this. Later, the gradient-descent algorithm (GDA, \cite{Madgwick2011}) is applied to the same problem as well. The performances are improved by levenberg-marquardt algorithm (LMA, \cite{Fourati2011}) and improved-GNA (IGNA, \cite{Tian2013}) in later literatures. Real-world applications have completely verified the feasibility, accuracy and computation speeds of such optimizers \cite{Valenti2016}.\\
\indent In previous optimizations, the attitude determination is treated as a nonlinear problem with the variable of quaternion, rotation vector, Euler angles and etc. This arouses a question: Is the problem convex or concave? As is known to us, the concave optimization suffers from local optimum in real practice. However, in aforementioned research, the convexity analysis has not been considered by the researchers. In fact, if the problem is concave then the performance of the optimizers would be significantly constrained for global searching. In a recent paper by S. Ahmed et al. \cite{Ahmed2012}, the authors declare that the attitude determination problem from vector observations is neither concave nor convex. In this paper, we give mathematical analysis aiming to show that the target problem is actually a convex one, leading to the robust insurance of current optimization solvers.\\
\indent This paper is arranged as follows: Section II addresses the problem background and our main results. Section III contains numerical examples. In Section IV, we present discussion and concluding remarks.

\section{Problem Background and Main Results}
A direction cosine matrix (DCM) relates a vector observation pair with
\begin{equation}
{{\mathbf{D}}^b} = {\mathbf{C}}{{\mathbf{D}}^r}
\end{equation}
where $\bf{C}$ denotes the DCM; ${{\mathbf{D}}^b}=(D_x^b,D_y^b,D_z^b)^T)$ and ${{\mathbf{D}}^r}=(D_x^r,D_y^r,D_z^r)^T)$ are the normalized vector observations from one sensor in the body frame $b$ and reference frame $r$ respectively. With several pairs of vector observations, the rotation matrix can be computed with the Wahba's problem that employs the following loss function
\begin{equation}
L({\mathbf{C}}) = {\sum\limits_{i = 1}^n {{a_i}\left\| {{\mathbf{D}}_i^b - {\mathbf{CD}}_i^r} \right\|} ^2}
\end{equation}
where $a_i$ is the positive weight of $i$-th sensor with $\sum\limits_{i = 1}^n {{a_i}}  = 1$. One can re-write this loss function into the system as follows
\begin{equation}
\left\{ {\begin{array}{*{20}{c}}
  {\sqrt {{a_1}} \left( {{\mathbf{D}}_1^b - {\mathbf{CD}}_1^r} \right) = {\mathbf{0}}} \\ 
  {\sqrt {{a_2}} \left( {{\mathbf{D}}_2^b - {\mathbf{CD}}_2^r} \right) = {\mathbf{0}}} \\ 
   \vdots  \\ 
  {\sqrt {{a_n}} \left( {{\mathbf{D}}_n^b - {\mathbf{CD}}_n^r} \right) = {\mathbf{0}}} 
\end{array}} \right.
\end{equation}
Optimization algorithms usually seek the minimum point of the Wahba's loss function by parameterizing the DCM with quaternion ${\bf{q}}=(q_0, q_1, q_2, q_3)^\top$, such that
\begin{equation} \label{optimization}
\mathop {\arg \min }\limits_{\left\| {\mathbf{q}} \right\| = 1} \sum\limits_{i = 1}^n {{a_i}{{\left\| {{\mathbf{D}}_i^b - {\mathbf{CD}}_i^r} \right\|}^2}} 
\end{equation}
A general solution i.e. the q-method solves the maximum eigenvalue and its associated eigenvector of the Davenport $\bf{K}$ matrix given as follows \cite{Yang2013}
\begin{equation}
{\mathbf{K}} = \left[ {\begin{array}{*{20}{c}}
  {{\cal{B}} + {{\cal{B}}^T} - tr({\cal{B}}){\bf{I}}}&{\mathbf{z}} \\ 
  {{{\mathbf{z}}^T}}&{tr({\cal{B}})} 
\end{array}} \right]
\end{equation}
where
\begin{equation}
\begin{gathered}
  {\cal{B}} = \sum\limits_{i = 1}^n {{a_i}{\mathbf{D}}_i^b{{\left( {{\mathbf{D}}_i^r} \right)}^T}}  \hfill \\
  {\mathbf{z}} = \sum\limits_{i = 1}^n {{a_i}{\mathbf{D}}_i^b \times {\mathbf{D}}_i^r}  \hfill \\ 
\end{gathered} 
\end{equation}
If the target multi-variate function is concave, there be local optimum setting up obstacles for global solving. In next sub-section, we are going to investigate the convexity of this optimization problem. 

\newcounter{mytempeqncnt}
\begin{figure*}[ht]
\setcounter{mytempeqncnt}{\value{equation}}
\setcounter{equation}{16}
\begin{equation}\label{P_C}
\begin{gathered}
  \frac{{{\partial ^2}{\mathbf{C}}}}{{\partial q_0^2}} = \left( {\begin{array}{*{20}{c}}
  2&0&0 \\ 
  0&2&0 \\ 
  0&0&2 
\end{array}} \right),\frac{{{\partial ^2}{\mathbf{C}}}}{{\partial {q_0}\partial {q_1}}} = \left( {\begin{array}{*{20}{c}}
  0&0&0 \\ 
  0&0&2 \\ 
  0&{ - 2}&0 
\end{array}} \right),\frac{{{\partial ^2}{\mathbf{C}}}}{{\partial {q_0}\partial {q_2}}} = \left( {\begin{array}{*{20}{c}}
  0&0&{ - 2} \\ 
  0&0&0 \\ 
  2&0&0 
\end{array}} \right),\frac{{{\partial ^2}{\mathbf{C}}}}{{\partial {q_0}\partial {q_3}}} = \left( {\begin{array}{*{20}{c}}
  0&2&0 \\ 
  { - 2}&0&0 \\ 
  0&0&0 
\end{array}} \right) \hfill \\
  \frac{{{\partial ^2}{\mathbf{C}}}}{{\partial {q_1}\partial {q_0}}} = \left( {\begin{array}{*{20}{c}}
  0&0&0 \\ 
  0&0&2 \\ 
  0&{ - 2}&0 
\end{array}} \right),\frac{{{\partial ^2}{\mathbf{C}}}}{{\partial q_1^2}} = \left( {\begin{array}{*{20}{c}}
  2&0&0 \\ 
  0&{ - 2}&0 \\ 
  0&0&{ - 2} 
\end{array}} \right),\frac{{{\partial ^2}{\mathbf{C}}}}{{\partial {q_1}\partial {q_2}}} = \left( {\begin{array}{*{20}{c}}
  0&2&0 \\ 
  2&0&0 \\ 
  0&0&0 
\end{array}} \right),\frac{{{\partial ^2}{\mathbf{C}}}}{{\partial {q_1}\partial {q_3}}} = \left( {\begin{array}{*{20}{c}}
  0&0&2 \\ 
  0&0&0 \\ 
  2&0&0 
\end{array}} \right) \hfill \\
  \frac{{{\partial ^2}{\mathbf{C}}}}{{\partial {q_2}\partial {q_0}}} = \left( {\begin{array}{*{20}{c}}
  0&0&{ - 2} \\ 
  0&0&0 \\ 
  2&0&0 
\end{array}} \right),\frac{{{\partial ^2}{\mathbf{C}}}}{{\partial {q_2}\partial {q_1}}} = \left( {\begin{array}{*{20}{c}}
  0&2&0 \\ 
  2&0&0 \\ 
  0&0&0 
\end{array}} \right),\frac{{{\partial ^2}{\mathbf{C}}}}{{\partial q_2^2}} = \left( {\begin{array}{*{20}{c}}
  { - 2}&0&0 \\ 
  0&2&0 \\ 
  0&0&{ - 2} 
\end{array}} \right),\frac{{{\partial ^2}{\mathbf{C}}}}{{\partial {q_2}\partial {q_3}}} = \left( {\begin{array}{*{20}{c}}
  0&0&0 \\ 
  0&0&2 \\ 
  0&2&0 
\end{array}} \right) \hfill \\
  \frac{{{\partial ^2}{\mathbf{C}}}}{{\partial {q_3}\partial {q_0}}} = \left( {\begin{array}{*{20}{c}}
  0&{2}&{0} \\ 
  {-2}&0&0 \\ 
  0&0&0 
\end{array}} \right),\frac{{{\partial ^2}{\mathbf{C}}}}{{\partial {q_3}\partial {q_1}}} = \left( {\begin{array}{*{20}{c}}
  0&0&2 \\ 
  0&0&0 \\ 
  2&0&0 
\end{array}} \right),\frac{{{\partial ^2}{\mathbf{C}}}}{{\partial {q_3}\partial {q_2}}} = \left( {\begin{array}{*{20}{c}}
  0&0&0 \\ 
  0&0&2 \\ 
  0&2&0 
\end{array}} \right),\frac{{{\partial ^2}{\mathbf{C}}}}{{\partial q_3^2}} = \left( {\begin{array}{*{20}{c}}
  {-2}&0&0 \\ 
  0&{-2}&0 \\ 
  0&0&2
\end{array}} \right) \hfill \\ 
\end{gathered} 
\end{equation}
\setcounter{equation}{\value{mytempeqncnt}}
\end{figure*}

\subsection{Single Vector Observation Pair}
Starting from a single vector observation pair, one can define the scalar loss  function as
\begin{equation}
{\mathcal{F}}_i({\mathbf{q}}) = {{\mathbf{e}}_i^T({\mathbf{q}}){{\mathbf{e}}_i}({\mathbf{q}})} 
\end{equation}
where ${{\mathbf{e}}_i}({\mathbf{q}}) = {\mathbf{D}}_i^b - {\mathbf{CD}}_i^r$ is the error vector function. Minimizing this target function can be achieved by GDA, GNA, LMA and etc. For instance, the LMA conduct optimization iteration by \cite{gill1981practical}
\begin{equation}
{{\mathbf{q}}_p} = {{\mathbf{q}}_{p - 1}} - \left( {{\mathbf{J}}_i^T{\mathbf{J}}_i^{} + \kappa {\mathbf{I}}} \right){\mathbf{J}}_i^T{{\mathbf{e}}_i}({{\mathbf{q}}_{p - 1}})
\end{equation}
where $p$ is the recursion index; $\kappa$ denotes a tiny positive number ensuring invertibility of matrix; ${{\mathbf{J}}_i^{}}$ stand for the Jacobian of ${{\mathbf{e}}_i}({{\mathbf{q}}_{p - 1}})$ with respect to ${{\mathbf{q}}_{p - 1}}$. Such optimization relies on the Hessian that determines whether there is local optimum or not. To study the convexity of the function ${\mathcal{F}}_i(\bf{q})$, we can simplify it into 
\begin{equation}
\begin{gathered}
  {\mathcal{F}}_i({\mathbf{q}}) = {\left( {{\mathbf{C}}{{\mathbf{D}}_i^r} - {{\mathbf{D}}_i^b}} \right)^T}\left( {{\mathbf{C}}{{\mathbf{D}}_i^r} - {{\mathbf{D}}_i^b}} \right) \hfill \\
   = {\left( {{{\mathbf{D}}_i^r}} \right)^T}{{\mathbf{C}}^T}{\mathbf{C}}{{\mathbf{D}}_i^r} + {\left( {{{\mathbf{D}}_i^b}} \right)^T}{{\mathbf{D}}_i^b} - {\left( {{{\mathbf{D}}_i^b}} \right)^T}{\mathbf{C}}{{\mathbf{D}}^r} - {\left( {{{\mathbf{D}}^r}} \right)^T}{{\mathbf{C}}^T}{{\mathbf{D}}_i^b} \hfill \\
   = 1 +  {\left( {q_0^2 + q_1^2 + q_2^2 + q_3^2} \right)^2} - {\left( {{{\mathbf{D}}_i^b}} \right)^T}{\mathbf{C}}{{\mathbf{D}}_i^r} - {\left( {{{\mathbf{D}}_i^r}} \right)^T}{{\mathbf{C}}^T}{{\mathbf{D}}_i^b} \hfill \\
   = 1 + {\left( {q_0^2 + q_1^2 + q_2^2 + q_3^2} \right)^2} - ({\cal{A}} + {\cal{A}}^T) \hfill \\ 
   = 1 + {\left( {q_0^2 + q_1^2 + q_2^2 + q_3^2} \right)^2} - 2{\cal{A}} \hfill \\
\end{gathered} 
\end{equation}
where we use
\begin{equation}
\begin{gathered}
  {\left( {{\mathbf{D}}_i^b} \right)^T}{\mathbf{D}}_i^b = {\left( {{\mathbf{D}}_i^r} \right)^T}{\mathbf{D}}_i^r = 1 \hfill \\
  {{\mathbf{C}}^T}{\mathbf{C}} = {\left( {q_0^2 + q_1^2 + q_2^2 + q_3^2} \right)^2} \hfill \\ 
\end{gathered} 
\end{equation}
and
\begin{equation}
{\cal{A}} = {\left( {{{\mathbf{D}}_i^b}} \right)^T}{\mathbf{C}}{{\mathbf{D}}_i^r}
\end{equation}
for simplification. The kernel problem is to deduce the convexity of function ${\cal{A}}$. Note that ${\left( {q_0^2 + q_1^2 + q_2^2 + q_3^2} \right)^2}={\left\| {\mathbf{q}} \right\|^4}$ is not simplified to 1 because of possible loss of normalization between successive optimization updates. The main thought of the proof is to show that the Hessian of ${\mathcal{F}}_i(\bf{q})$ belongs to positive semidefinite matrices \cite{boyd2004convex}.

\begin{figure*}[ht]
\setcounter{mytempeqncnt}{\value{equation}}
\setcounter{equation}{17}
\begin{equation}\label{QA}
{\mathbf{Q}}_{\cal{A}}^{} = \left[ {\begin{array}{*{20}{c}}
  {\frac{{\left( {D_x^b - D_x^r} \right)\left( {D_z^b - D_z^r} \right)}}{N}}&{ - \frac{{\left( {D_x^b - D_x^r} \right)\left( {D_y^b - D_y^r} \right)}}{N}}&{\frac{{\left( {D_x^b + D_x^r} \right)\left( {D_z^b + D_z^r} \right)}}{N}}&{ - \frac{{\left( {D_x^b + D_x^r} \right)\left( {D_y^b + D_y^r} \right)}}{N}} \\ 
  {\frac{{\left( {D_x^b - D_x^r} \right)\left( {D_y^b + D_y^r} \right)}}{N}}&{\frac{{\left( {D_x^b - D_x^r} \right)\left( {D_z^b + D_z^r} \right)}}{N}}&{\frac{{\left( {D_y^b - D_y^r} \right)\left( {D_x^b + D_x^r} \right)}}{N}}&{\frac{{\left( {D_x^b - D_x^r} \right)\left( {D_z^b - D_z^r} \right)}}{N}} \\ 
  1&0&1&0 \\ 
  0&1&0&1 
\end{array}} \right]
\end{equation}
\setcounter{equation}{\value{mytempeqncnt}}
\end{figure*}


Taking the Hessian of ${\mathcal{F}}_i({\bf{q}})$, we obtain
\begin{equation}\label{H_init}
{{\mathbf{H}}_{{{\mathcal{F}}_i}}} = \left( {\begin{array}{*{20}{c}}
  {\frac{{{\partial ^2}{\mathcal{F}}_i}}{{\partial q_0^2}}}&{\frac{{{\partial ^2}{\mathcal{F}}_i}}{{\partial {q_0}\partial {q_1}}}}&{\frac{{{\partial ^2}{\mathcal{F}}_i}}{{\partial {q_0}\partial {q_2}}}}&{\frac{{{\partial ^2}{\mathcal{F}}_i}}{{\partial {q_0}\partial {q_3}}}} \\ 
  {\frac{{{\partial ^2}{\mathcal{F}}_i}}{{\partial {q_1}\partial {q_0}}}}&{\frac{{{\partial ^2}{\mathcal{F}}_i}}{{\partial q_1^2}}}&{\frac{{{\partial ^2}{\mathcal{F}}_i}}{{\partial {q_1}\partial {q_2}}}}&{\frac{{{\partial ^2}{\mathcal{F}}_i}}{{\partial {q_1}\partial {q_3}}}} \\ 
  {\frac{{{\partial ^2}{\mathcal{F}}_i}}{{\partial {q_2}\partial {q_0}}}}&{\frac{{{\partial ^2}{\mathcal{F}}_i}}{{\partial {q_2}\partial {q_1}}}}&{\frac{{{\partial ^2}{\mathcal{F}}_i}}{{\partial q_2^2}}}&{\frac{{{\partial ^2}{\mathcal{F}}_i}}{{\partial {q_2}\partial {q_3}}}} \\ 
  {\frac{{{\partial ^2}{\mathcal{F}}_i}}{{\partial {q_3}\partial {q_0}}}}&{\frac{{{\partial ^2}{\mathcal{F}}_i}}{{\partial {q_3}\partial {q_1}}}}&{\frac{{{\partial ^2}{\mathcal{F}}_i}}{{\partial {q_3}\partial {q_2}}}}&{\frac{{{\partial ^2}{\mathcal{F}}_i}}{{\partial q_3^2}}} 
\end{array}} \right)
\end{equation}
in which
\begin{equation}
\frac{{\partial {\cal{A}}}}{{\partial {q_k}}} = {\left( {{\mathbf{D}}_i^b} \right)^T}\frac{{\partial {\mathbf{C}}}}{{\partial {q_k}}}{\mathbf{D}}_i^r,k = 0,1,2,3
\end{equation}
where $k=0,1,2,3$ are the quaternion indices and the derivatives of $\bf{C}$ can be computed by
\begin{equation}
\begin{gathered}
  \frac{{\partial {\mathbf{C}}}}{{\partial {q_0}}} = 2\left( {\begin{array}{*{20}{c}}
  {{q_0}}&{{q_3}}&{ - {q_2}} \\ 
  { - {q_3}}&{{q_0}}&{{q_1}} \\ 
  {{q_2}}&{ - {q_1}}&{{q_0}} 
\end{array}} \right) \hfill \\
  \frac{{\partial {\mathbf{C}}}}{{\partial {q_1}}} = 2\left( {\begin{array}{*{20}{c}}
  {{q_1}}&{{q_2}}&{{q_3}} \\ 
  {{q_2}}&{ - {q_1}}&{{q_0}} \\ 
  {{q_3}}&{ - {q_0}}&{ - {q_1}} 
\end{array}} \right) \hfill \\
  \frac{{\partial {\mathbf{C}}}}{{\partial {q_2}}} = 2\left( {\begin{array}{*{20}{c}}
  { - {q_2}}&{{q_1}}&{ - {q_0}} \\ 
  {{q_1}}&{{q_2}}&{{q_3}} \\ 
  {{q_0}}&{{q_3}}&{ - {q_2}} 
\end{array}} \right) \hfill \\
  \frac{{\partial {\mathbf{C}}}}{{\partial {q_3}}} = 2\left( {\begin{array}{*{20}{c}}
  { - {q_3}}&{{q_0}}&{{q_1}} \\ 
  { - {q_0}}&{ - {q_3}}&{{q_2}} \\ 
  {{q_1}}&{{q_2}}&{{q_3}} 
\end{array}} \right) \hfill \\ 
\end{gathered}
\end{equation}
The above equations lead to further computations in (\ref{P_C}). Then we obtain
\begin{equation}
\frac{{{\partial ^2}{\cal{A}}}}{{\partial {q_k}\partial {q_j}}} = {\left( {{{\mathbf{D}}_i^b}} \right)^T}\frac{{{\partial ^2}{\mathbf{C}}}}{{\partial {q_k}\partial {q_j}}}{{\mathbf{D}}_i^r},\left\{ {\begin{array}{*{20}{c}}
  {k = 0,1,2,3} \\ 
  {j = 0,1,2,3} 
\end{array}} \right.
\end{equation}
From the results of second-order derivative of $\bf{C}$ we can observe that
\begin{equation}
\frac{{\partial {\mathbf{C}}}}{{\partial {q_k}\partial {q_j}}} = \frac{{\partial {\mathbf{C}}}}{{\partial {q_j}\partial {q_k}}},\left\{ {\begin{array}{*{20}{c}}
  {k = 0,1,2,3} \\ 
  {j = 0,1,2,3} 
\end{array}} \right.
\end{equation}
This leads to the Hessian of $A$ i.e. ${\bf{H}}_{{\cal{A}}}$ being a symmetric matrix, such that
\setcounter{equation}{18}
\begin{equation}
\begin{gathered}
  {H_{{\cal{A}},11}} = 2\left( {D_x^bD_x^r + D_y^bD_y^r + D_z^bD_z^r} \right) \hfill \\
  {H_{{\cal{A}},12}} = 2\left( {D_z^bD_y^r - D_y^bD_z^r} \right) \hfill \\
  {H_{{\cal{A}},13}} = 2\left( {D_x^bD_z^r - D_z^bD_x^r} \right) \hfill \\
  {H_{{\cal{A}},14}} = 2\left( {D_y^bD_x^r - D_x^bD_y^r} \right) \hfill \\
  {H_{{\cal{A}},22}} = 2\left( { - D_x^bD_x^r + D_y^bD_y^r + D_z^bD_z^r} \right) \hfill \\
  {H_{{\cal{A}},23}} =  - 2\left( {D_y^bD_x^r + D_x^bD_y^r} \right) \hfill \\
  {H_{{\cal{A}},24}} =  - 2\left( {D_z^bD_x^r + D_x^bD_z^r} \right) \hfill \\
  {H_{{\cal{A}},33}} = 2\left( {D_x^bD_x^r - D_y^bD_y^r + D_z^bD_z^r} \right) \hfill \\
  {H_{{\cal{A}},34}} =  - 2\left( {D_z^bD_y^r + D_y^bD_z^r} \right) \hfill \\
  {H_{{\cal{A}},44}} = 2\left( {D_x^bD_x^r + D_y^bD_y^r - D_z^bD_z^r} \right) \hfill \\ 
\end{gathered} 
\end{equation}
where $H_{{\cal{A}}, jk}$ is the entry of ${\bf{H}}_{\cal{A}}$ in the $j$-th row and $k$-th column. The Hessian of ${\left( {q_0^2 + q_1^2 + q_2^2 + q_3^2} \right)^2}$ can easily be computed by
\begin{equation}
{{\bf{H}}_{{{\left\| {\bf{q}} \right\|}^4}}} = 4\left( {q_0^2 + q_1^2 + q_2^2 + q_3^2} \right){\bf{I}} + 8{\bf{q}}{{\bf{q}}^T}
\end{equation}
Hence the final Hessian of $F_i(\bf{q})$ takes the following form
\begin{equation}
{{\mathbf{H}}_{{\cal{F}}_i}} = 4{\left\| {\mathbf{q}} \right\|^2}{\mathbf{I}} + 8{\mathbf{q}}{{\mathbf{q}}^T} - 2{{\mathbf{H}}_{\cal{A}}}
\end{equation}
Notice that ${\bf{H}}_{\cal{A}}$ has the eigenvalue decomposition of \cite{Wu2017}
\begin{equation}
{{\mathbf{H}}_{\cal{A}}} = 2{\mathbf{Q}}_{\cal{A}}^{}{{\bf{\Sigma}} _{\cal{A}}}{\mathbf{Q}}_{\cal{A}}^{ - 1}
\end{equation}
where
\begin{equation}
{{\bf{\Sigma}} _{\cal{A}}} = diag(1,1, - 1, - 1)
\end{equation}
while ${\mathbf{Q}}_{\cal{A}}^{}$ is given in (\ref{QA}) in which
\begin{equation}
N = {\left( {D_y^b} \right)^2} + {\left( {D_z^b} \right)^2} - {\left( {D_y^r} \right)^2} - {\left( {D_y^r} \right)^2}
\end{equation}
Therefore we can see that
$4{\left\| {\mathbf{q}} \right\|^2}{\mathbf{I}} - 2{{\mathbf{H}}_{\cal{A}}}$ has the eigenvalues of
\begin{equation}
{\lambda _{4{{\left\| {\mathbf{q}} \right\|}^2}{\mathbf{I}} - 2{{\mathbf{H}}_{\cal{A}}}}} = \left\{ {\begin{array}{*{20}{c}}
  {4{{\left\| {\mathbf{q}} \right\|}^2} + 4} \\ 
  {4{{\left\| {\mathbf{q}} \right\|}^2} + 4} \\ 
  {4{{\left\| {\mathbf{q}} \right\|}^2} - 4} \\ 
  {4{{\left\| {\mathbf{q}} \right\|}^2} - 4} 
\end{array}} \right.
\end{equation}
From another aspect, ${\mathbf{q}}{{\mathbf{q}}^T}$ is a matrix with rank 1 and all non-negative eigenvalues, such that
\begin{equation}
{\mathbf{q}}{{\mathbf{q}}^T} = {{\mathbf{Q}}_{{\mathbf{q}}{{\mathbf{q}}^T}}}{{\bf{\Sigma}} _{{\mathbf{q}}{{\mathbf{q}}^T}}}{\mathbf{Q}}_{{\mathbf{q}}{{\mathbf{q}}^T}}^{ - 1}
\end{equation}
where
\begin{equation}
\begin{gathered}
  {{\bf{\Sigma}} _{{\mathbf{q}}{{\mathbf{q}}^T}}} = diag\left( {0,0,0,q_0^2 + q_1^2 + q_2^2 + q_3^2} \right) \hfill \\
  {{\mathbf{Q}}_{{\mathbf{q}}{{\mathbf{q}}^T}}} = \left( {\begin{array}{*{20}{c}}
  { - \frac{{{q_3}}}{{{q_0}}}}&0&0&1 \\ 
  { - \frac{{{q_2}}}{{{q_0}}}}&0&1&0 \\ 
  { - \frac{{{q_1}}}{{{q_0}}}}&1&0&0 \\ 
  {\frac{{{q_0}}}{{{q_3}}}}&{\frac{{{q_1}}}{{{q_3}}}}&{\frac{{{q_2}}}{{{q_3}}}}&1 
\end{array}} \right) \hfill \\ 
\end{gathered} 
\end{equation}
Therefore we have
\begin{equation}
\begin{gathered}
  rank({{\mathbf{H}}_{{\cal{F}}_i}}) = rank(4{{{\left\| {\mathbf{q}} \right\|}^2}}{\mathbf{I}} - 2{{\mathbf{H}}_{\cal{A}}} + 8{\mathbf{q}}{{\mathbf{q}}^T}) \leqslant  \hfill \\
  \begin{array}{*{20}{c}}
  {}&{}&{}&{} 
\end{array}rank(4{{{\left\| {\mathbf{q}} \right\|}^2}}{\mathbf{I}} - 2{{\mathbf{H}}_{\cal{A}}}) + rank(8{\mathbf{q}}{{\mathbf{q}}^T})\hfill \\ 
\end{gathered} 
\end{equation}
The eigenvalues of ${{\mathbf{H}}_{{\cal{F}}_i}}$ satisfies
\begin{equation}
\begin{gathered}
  \min ({\lambda _{4{{{\left\| {\mathbf{q}} \right\|}^2}}{\mathbf{I}} - 2{{\mathbf{H}}_{\cal{A}}}}}) + \min ({\lambda _{8{\mathbf{q}}{{\mathbf{q}}^T}}}) \leqslant {\lambda _{\mathbf{H}}} \hfill \\
  \begin{array}{*{20}{c}}
  {}&{}&{}&{}&{} 
\end{array} \leqslant \max ({\lambda _{4{{{\left\| {\mathbf{q}} \right\|}^2}}{\mathbf{I}} - 2{{\mathbf{H}}_{\cal{A}}}}}) + \max ({\lambda _{8{\mathbf{q}}{{\mathbf{q}}^T}}}) \hfill \\ 
\end{gathered} 
\end{equation}
yielding
\begin{equation}
4{\left\| {\mathbf{q}} \right\|^2} - 4 \leqslant {\lambda _{{{\mathbf{H}}_i}}} \leqslant 12{\left\| {\mathbf{q}} \right\|^2} + 4
\end{equation}
That is to say, ${{\mathbf{H}}_{{\cal{F}}_i}}$ is currently indefinite. However, when conducting in the optimization ensuring that the quaternion is always normalized in last step, we would obtain $\left\| {\mathbf{q}} \right\| = 1$. In fact, for all $\bf{q}$ owning $\left\| {\mathbf{q}} \right\| \geqslant 1$, ${{\mathbf{H}}_{{\cal{F}}_i}}$ is a positive semidefinite symmetric matrix with rank 3 or 4. As in all literatures, normalization of quaternion always takes place, then it is ensured that the optimization is convex \cite{boyd2004convex}. 

\subsection{Multi-Vector Case}
Defining 
\begin{equation}
{\bf{e}}({\mathbf{q}}) = \left[ {\begin{array}{*{20}{c}}
  {\sqrt {{a_1}} \left( {{\mathbf{D}}_1^b - {\mathbf{CD}}_1^r} \right)} \\ 
  {\sqrt {{a_2}} \left( {{\mathbf{D}}_2^b - {\mathbf{CD}}_2^r} \right)} \\ 
   \vdots  \\ 
  {\sqrt {{a_n}} \left( {{\mathbf{D}}_n^b - {\mathbf{CD}}_n^r} \right)} 
\end{array}} \right]
\end{equation}
, one can easily find out that the target loss function defined by
\begin{equation}
{\cal{F}}({\mathbf{q}}) = {{\mathbf{e}}^T}({\mathbf{q}}){\mathbf{e}}({\mathbf{q}})
\end{equation}
meets
\begin{equation}
{\cal{F}}({\mathbf{q}}) = \sum\limits_{i = 1}^n {{a_i}{{\cal{F}}_i}({\mathbf{q}}) = } \sum\limits_{i = 1}^n {{a_i}{{\left\| {{\mathbf{D}}_i^b - {\mathbf{CD}}_i^r} \right\|}^2}} 
\end{equation}
Its Hessian $\bf{H}$
\begin{equation}
{\mathbf{H}} = \sum\limits_{i = 1}^n {{a_i}{{\mathbf{H}}_{{{\cal{F}}_i}}}} 
\end{equation}
has the eigenvalue inequality of
\begin{equation}
\begin{gathered}
  0 \leqslant {\lambda _{\mathbf{H}}} \leqslant \sum\limits_{i = 1}^n {{a_i}\left[ {4 + 12\left( {q_0^2 + q_1^2 + q_2^2 + q_3^2} \right)} \right]}  \hfill \\
  \begin{array}{*{20}{c}}
  {}&{}&{}&{}&{} 
\end{array} = 4 + 12\left( {q_0^2 + q_1^2 + q_2^2 + q_3^2} \right) \hfill \\ 
\end{gathered} 
\end{equation}
which proves the convexity of the attitude optimization from multi-vector observations.

\section{Numerical Example}
Assume that we obtain the following single vector observation pair from a vector sensor
\begin{equation}
\begin{gathered}
  {{\mathbf{D}}^b} = \left( {\begin{array}{*{20}{c}}
  {{ - 0.712824827533344}} \\ 
  {{ - 0.225772381096068}} \\ 
  {{0.664008732763561}} 
\end{array}} \right) \hfill \\
  {{\mathbf{D}}^r} = \left( {\begin{array}{*{20}{c}}
  {{- 0.037453665434217}} \\ 
  {{0.500499809534146}} \\ 
  {{ - 0.864926102971707}} 
\end{array}} \right) \hfill \\ 
\end{gathered} 
\end{equation}
When we perform optimization based on last unnormalized updated quaternion such as
\begin{equation}
{\mathbf{q}} = \left( {\begin{array}{*{20}{c}}
  {{0.420683700201250}} \\ 
  {{0.400737998146962}} \\ 
  {{0.095142157864169}} \\ 
  {{0.496684391636530}} 
\end{array}} \right)
\end{equation}
with $\left\| {\mathbf{q}} \right\| = {0.770268222031943} < 1$, the Hessian's eigenvalues can be computed by
\begin{equation}
{\lambda _{\mathbf{H}}} = \left\{ {\begin{array}{*{20}{c}}
  {{9.761874553883407}} \\ 
  {{6.373252535488999}} \\ 
  { - {0.268864411927401}} \\ 
  { - {1.626747464510996}} 
\end{array}} \right.
\end{equation}
indicating that the optimization is non-convex nor concave. However, with normalized quaternion of 
\begin{equation}
{\mathbf{q}} = \left( {\begin{array}{*{20}{c}}
  {{0.118759061535262}} \\ 
  { - {0.346543560044311}} \\ 
  { - {0.817997262250335}} \\ 
  {{0.443491065576337}} 
\end{array}} \right)
\end{equation}
we can compute the Hessian's eigenvalues by
\begin{equation}
{\lambda _{\mathbf{H}}} = \left\{ {\begin{array}{*{20}{c}}
  {{14.677631236006699}} \\ 
  {{7.999999999999996}} \\ 
  {{1.322368763993306}} \\ 
  {{0.000000000000000}} 
\end{array}} \right.
\end{equation}
which reflects $\bf{H}$ here is positive semidefinite. While with
\begin{equation}
{\mathbf{q}} = \left( {\begin{array}{*{20}{c}}
  {{ - 0.353622599299341}} \\ 
  {{0.046434526687823}} \\ 
  {{0.046434526687823}} \\ 
  {{- 1.550514474779561}} 
\end{array}} \right)
\end{equation}
in whch $\left\| {\mathbf{q}} \right\| = {1.777657443309303 > 1}$, we have
\begin{equation}
{\lambda _{\mathbf{H}}} = \left\{ {\begin{array}{*{20}{c}}
  {{39.127609299877825}} \\ 
  {{16.640263943011881}} \\ 
  {{11.433446472169676}} \\ 
  {{8.640263943011886}} 
\end{array}} \right.
\end{equation}
which verifies the former derived results that the current $\bf{H}$ owns positive definiteness. Therefore, we verify that that the attitude determination problem in (\ref{optimization}) is always convex provided that the last-step quaternion is normalized.

\section{Conclusion}
In this paper, the optimization framework of the attitude determination from vector observations is revisited. Some closed-form results are derived showing the identities of the Hessian to the optimization. By eigenvalue analysis, it is found out that the original problem is sometimes concave but is rigorously convex after adding a quaternion normalization in advance. As such commitment is very common in real engineering practice ensuring unitary quaternion norm, the target optimization can be regarded as a fully convex one. Numerical examples containing simulated cases verify the derived results. It is fully proven in this paper that the previous derivative-based optimization techniques are robust in the case of convexity. And we hope that this contribution would benefit related research in the future.


%

%

\section*{Acknowledgment}
This work was supported by National Natural Science Foundation of China under the grant of No. 41604025 and was also supported by State Key Laboratory of Geodesy and Earth's Dynamics (Institute of Geodesy and Geophysics, Chinese Academy of Sciences) Grant No. SKLGED2018-3-2-E.

\ifCLASSOPTIONcaptionsoff
  \newpage
\fi

\end{document}